\documentclass[12pt]{article}
\usepackage{epsfig}

\renewcommand{\theequation}{\arabic{section}.\arabic{equation}}

\def\beq{\begin{equation}}
\def\eeq{\end{equation}}
\def\bea{\begin{eqnarray}}
\def\eea{\end{eqnarray}}

\def\gev{\, {\rm GeV}}

\newcommand{\gsim}{\lower.7ex\hbox{$\;\stackrel{\textstyle>}{\sim}\;$}}
\newcommand{\lsim}{\lower.7ex\hbox{$\;\stackrel{\textstyle<}{\sim}\;$}}

\def\stilde{\widetilde}

\newcommand{\newc}{\newcommand}
\newc{\Nc}{N_{c}}
\newc{\CG}{C_G}
\newc{\gp}{g'}
\newc{\stopi}{\stilde t_i}
\newc{\sboti}{\stilde b_i}
\newc{\staui}{\stilde \tau_i}
\newc{\stopj}{\stilde t_j}
\newc{\sbotj}{\stilde b_j}
\newc{\stauj}{\stilde \tau_j}
\newc{\stopI}{\stilde t_1}
\newc{\stopII}{\stilde t_2}
\newc{\sbotI}{\stilde b_1}
\newc{\sbotII}{\stilde b_2}
\newc{\stauI}{\stilde \tau_1}
\newc{\stauII}{\stilde \tau_2}
\newc{\sstop}{s_{t}}
\newc{\cstop}{c_{t}}
\newc{\ssbot}{s_{b}}
\newc{\csbot}{c_{b}}
\newc{\sstau}{s_{\tau}}
\newc{\cstau}{c_{\tau}}
\newc{\Sstop}{s_{2t}}
\newc{\Cstop}{c_{2t}}
\newc{\Ssbot}{s_{2b}}
\newc{\Csbot}{c_{2b}}
\newc{\Sstau}{s_{2\tau}}
\newc{\Cstau}{c_{2\tau}}
\newc{\salpha}{s_\alpha}
\newc{\calpha}{c_\alpha}
\newc{\Calpha}{c_{2\alpha}}
\newc{\Salpha}{s_{2\alpha}}
\newc{\sbetapm}{s_{\beta_\pm}}
\newc{\cbetapm}{c_{\beta_\pm}}
\newc{\Sbetapm}{s_{2 \beta_\pm}}
\newc{\Cbetapm}{c_{2 \beta_\pm}}
\newc{\sbetaO}{s_{\beta_0}}
\newc{\cbetaO}{c_{\beta_0}}
\newc{\SbetaO}{s_{2 \beta_0}}
\newc{\CbetaO}{c_{2 \beta_0}}
\newc{\vu}{v_u}
\newc{\vd}{v_d}
\newc{\seL}{\stilde e_L}
\newc{\smuL}{\stilde \mu_L}
\newc{\seR}{\stilde e_R}
\newc{\smuR}{\stilde \mu_R}
\newc{\suL}{\stilde u_L}
\newc{\sdL}{\stilde d_L}
\newc{\suR}{\stilde u_R}
\newc{\sdR}{\stilde d_R}
\newc{\scL}{\stilde c_L}
\newc{\ssL}{\stilde s_L}
\newc{\scR}{\stilde c_R}
\newc{\ssR}{\stilde s_R}
\newc{\snue}{\stilde \nu_e}
\newc{\snumu}{\stilde \nu_\mu}
\newc{\snutau}{\stilde \nu_\tau}
\newc{\Gpm}{G^\pm}
\newc{\Hpm}{H^\pm}
\newc{\FFbS}{\overline{FF}S}
\newc{\FFbV}{\overline{FF}V}
\newc{\FSS}{F_{SS}}
\newc{\FSSS}{F_{SSS}}
\newc{\FFFS}{F_{FFS}}
\newc{\FFFbS}{F_{\overline{FF}S}}
\newc{\FSSV}{F_{SSV}}
\newc{\FVS}{F_{VS}}
\newc{\FVVS}{F_{VVS}}
\newc{\FFFV}{F_{FFV}}
\newc{\FFFbV}{F_{\overline{FF}V}}
\newc{\Fgauge}{F_{\rm gauge}}
\newc{\DRbarprime}{$\overline{\rm DR}'$ }
\newc{\DRbar}{$\overline{\rm DR}$ }
\newc{\MSbar}{$\overline{\rm MS}$ }
\newc{\Yu}{{\bf Y}_u}
\newc{\Yd}{{\bf Y}_d}
\newc{\Ye}{{\bf Y}_e}
\newc{\Au}{{\bf a}_u}
\newc{\Ad}{{\bf a}_d}
\newc{\Ae}{{\bf a}_e}
\newc{\bm}{{\bf m}}

\newc{\rwino}{r_{\tilde W}}
\newc{\rmu}{r_{\tilde H}}
\newc{\ra}{r_A}

\newcommand{\nc}{\newcommand}

\nc{\beaa}{\begin{eqnarray*}}   \nc{\eeaa}{\end{eqnarray*}}

\nc{\baa}{\begin{array}}      \nc{\eaa}{\end{array}}
\def\bit{\begin{itemize}}    
\def\eit{\end{itemize}}
\nc{\ben}{\begin{enumerate}}  \nc{\een}{\end{enumerate}}
\nc{\bce}{\begin{center}}     \nc{\ece}{\end{center}}
\def\bed{\begin{description}}
\def\eed{\end{description}}

\def\eps{\epsilon}
\def\non{\nonumber}

\def\k1slash{k_1\hspace{-10.5pt}/\ \ }

\def\msqi{M_{\tilde{q}_i}}

\def\mgss{M_{\tilde{g}}}
\def\mchi{M_{\tilde{\chi}}}

\begin{document}

\setlength{\baselineskip}{0.25in}


\begin{titlepage}
\noindent
\begin{flushright}
MCTP-05-46
\end{flushright}
\vspace{1cm}

\begin{center}
  \begin{Large}
    \begin{bf}
Gluino decays with heavier scalar superpartners
    \end{bf}
  \end{Large}
\end{center}
\vspace{0.2cm}
\begin{center}
\begin{large}
Manuel Toharia and James D. Wells \\
\end{large}
  \vspace{0.3cm}
  \begin{it}
\vspace{0.1cm}
Michigan Center for Theoretical Physics (MCTP) \\
        ~~University of Michigan, Ann Arbor, MI 48109-1120, USA \\
\vspace{0.1cm}
\end{it}

\end{center}

\begin{abstract}

We compute gluino decay widths in supersymmetric theories
with arbitrary flavor and CP violation angles.  Our emphasis
is on theories with scalar superpartner masses heavier than
the gluino such that tree-level two-body decays are not allowed,
which is relevant, for example, in split supersymmetry.
We compute gluino decay branching fractions in several specific
examples and show that it is plausible that the only 
accessible signal
of supersymmetry at the LHC could be four top quarks plus
missing energy.  We show another example where the only 
accessible signal
for supersymmetry is two gluon jets plus missing energy.

\end{abstract}

\vspace{1cm}

\begin{flushleft}
hep-ph/0503175 \\
\end{flushleft}

\end{titlepage}

\setcounter{footnote}{1}
\setcounter{page}{2}
\setcounter{figure}{0}
\setcounter{table}{0}


\section{Introduction}
\bigskip

A large class of supersymmetry breaking ideas predict gauginos
with less mass than scalar superpartners of the standard model 
fermions~\cite{amsb}.
It has been emphasized recently that a good-sized separation between
the gauginos and scalars could help solve some of supersymmetry's 
problems by suppressing flavor and CP violating effects, 
while maintaining its good features such as gauge coupling unification
and dark matter. 

Within these general ideas of extraordinarily heavy
scalar particles~\cite{split susy,split susy 2} or near PeV-scale
scalars~\cite{pev susy},
there is no reason to expect the superpartner 
flavor angles to be diagonal, aligned or symmetrized in any way.  Furthermore,
there is no reason to expect the CP violating phases of soft terms
to be significantly suppressed. In short, anything goes with these angles, and 
computation of the important phenomenological implications of these
models must take into account this freedom.

Perhaps the most important phenomenological handle on split supersymmetry
from a collider physics point of view is the production and decay of
gluinos~\cite{pev susy}. 
So far in the literature there has been no complete calculation
of the gluino decay widths with arbitrary flavor angles and CP violating
phases.  In this article our first goal is to present a 
complete calculation of the gluino decays with arbitrary flavor 
angles and CP violating phases. Furthermore,
from the structure of the amplitudes that we present, the reader can
quite trivially compute gluino decays in theories where there are more
neutralinos and/or charginos than the MSSM requires, as would be expected
for example in a theory with an extra $U(1)$ gauge group factor.  

Our second goal is to compute the gluino decay
branching fractions in some interesting simplified models that can
show the rich variety of possibilities that gluino decays could present
to us at the LHC. For example, it is plausible that the only accessible signal
of supersymmetry at the LHC could be four top quarks plus
missing energy.  We show another example where the only 
accessible signal
for supersymmetry is two gluon jets plus missing energy.
Many other possible phenomenologies exist, which we illustrate below.

\section{Gluino decay}
\setcounter{equation}{0}
\setcounter{footnote}{1}

\begin{figure}[t]
\centering
\includegraphics[width=13cm]{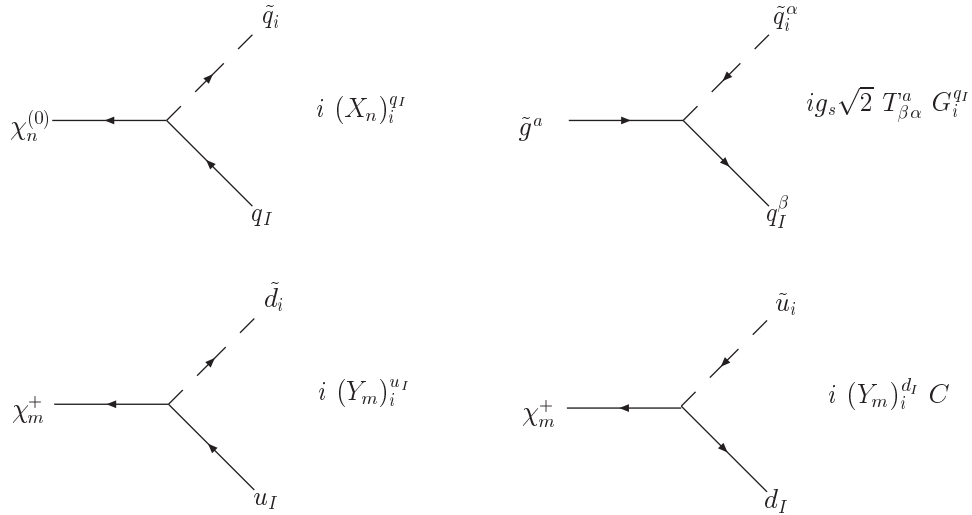}
\caption{Generic Feynman rules for ``-ino''$-$quark$-$squark interactions.}
\label{feynrul}
\vspace{.3cm}
\end{figure}

Gluino decays when the squarks are heavier than the gluino
itself have been
studied in the past in some models of supersymmetry. In this case gluinos can
undergo a three body decay into two quarks and a neutralino or chargino 
\cite{3bod}, or decay radiatively
into a gluon and a neutralino \cite{2bod,Haber-Wyler}.
With the usual universality conditions,
third family squarks and sleptons can have a sizable mixing, and such
effects have also been explicitly included in the literature \cite{Bartlb}.

Nevertheless, it is useful to compute the decay width
formulae in a more general (and compact) way as we discussed 
in the introduction.
We include in this section the complete formulae for gluino decays
in the case where squarks are heavier than the gluino. 
The formulae are left in terms of 
the general couplings between quarks, squarks and ``-inos'' 
(gluinos, neutralinos and charginos) shown in Figure~\ref{feynrul}.
Since these are generic couplings, one can add flavor
mixing among squarks (unconstrained when the squarks are heavier than
about $10^5$ GeV), CP phases, or include extra neutralinos in the
spectrum.
One simply has to explicitly compute the Feynman rules of
Figure~\ref{feynrul} for the desired scenario and introduce them into
the formulae.
After a phase space integration for the three body decays or a one-loop
integration for the two body decays (both easily done numerically),
one can get precise gluino decay widths for many extensions of
the usual MSSM scenarios.

{\it Definitions, conventions and notation}

There are three basic types of ``-ino''$-$quark$-$squark interactions which are
shown in Figure~\ref{feynrul}. We define them by $(X_n)^{q_I}_i$, $\
G^{q_I}_i$, $\ (Y_m)^{u_I}_i$ and
$(Y_m)^{d_I}_i$.

We also need to define the ``tilded'' couplings in terms of the Dirac
matrix $\gamma_0$,
\beq
\widetilde{G}^{q_I}_i=\gamma_0 G^{q_I}_i\gamma_0 \ , \ \ \ \ \
\widetilde{X}^{q_I}_i=\gamma_0 X_i^{q_I} \gamma_0\ , \ \ \ \ \
\widetilde{Y}^{u_I}_i=\gamma_0 Y_i^{u_I} \gamma_0\ \ \  {\rm and}\ \ \ 
\widetilde{Y}^{d_I}_i=\gamma_0 Y_i^{d_I} \gamma_0\
\eeq

The indices that appear in the formulae are defined as follows:
the index $i$ runs through the six squarks of both up and down
  type $(i=1,...\, ,6)$, whereas the index $I$ runs through the 3 quarks
  of both up and down type $(I=1,2,3)$.
The index $q$ denotes the type of family in question, $up$ or $down$ $(q=u,d)$.
Finally $n$ and $m$ refer to the neutralino and chargino physical states
respectively. Since each chargino and neutralino are specific external particles we
will generally omit their index number inside the formulae.

{\it Decay channels and widths}

We now present the complete two-body (radiative) and three-body decay
widths of the gluino (in the heavy squark scenario).

\begin{figure}[t]
\centering
\includegraphics[width=15cm]{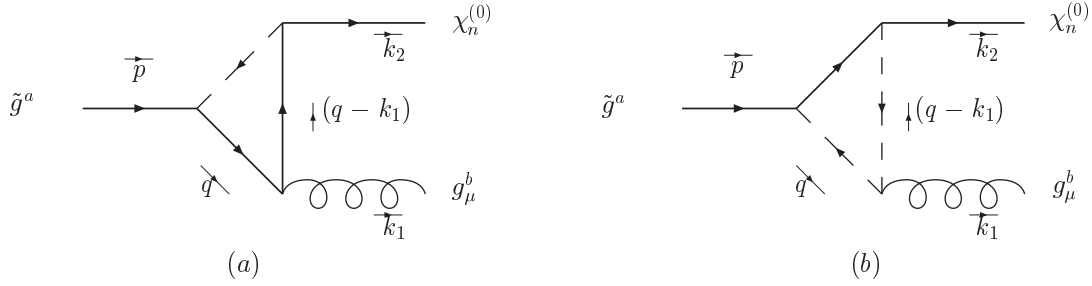}
\vspace{-.3cm}
\caption{Diagrams contributing to the one-loop radiative gluino decay
  in MSSM. Due to the majorana nature of gluino and neutralino, two more
  diagrams contribute, but they differ only from a) and b)
  by having opposite fermion lines (the flow of charge inside the loop
  changes direction)}
\label{diags}
\vspace{.3cm}
\end{figure}

\bit
\item $\tilde{g}\to g \chi^o_{\rm n} $

The decay width for this process (see Figure~\ref{diags}) is
\bea
\Gamma(\tilde{g}\to g \chi^o_{\rm n})={1\over 8 \pi } {1\over32^2
  \pi^4} {g^2} {g_s^4}\ \left({\mgss^2-\mchi^2\over \mgss}\right)^3\ {1\over4}{{\rm Tr}\left[{\bf \left(C_{n}\right)^\dagger 
\left(C_{n}\right)}   \right]} 
\eea
where the trace is computed in Dirac Space, given the chiral structure
of the coupling matrices $G^q_i$ and $X^q_i$, which are part of the
effective coupling ${\bf C_{n}}$, defined by\footnote{We have checked
  that this formula agrees with~\cite{Haber-Wyler} (by properly
  converting their neutralino decay into gluino decay) and with Baer et
al. in~\cite{2bod} in the limit of no mixing.}
\bea
{g\over\sqrt{2}}{\bf\left(C_{n}\right)}\!\!&=&\!\!\sum_{i,q_I}\left(   M_{\tilde{g}}\
   \left( X^{q_I}_i\widetilde{G}^{q_I}_i-\eps_n \eps_g \widetilde{X}^{{q_I}{{}^\dagger}}_i G^{{q_I}{{}^\dagger}}_i\right)
  \left(C_{23}+C'_{23}+2C_{12}\right)\Big|_{M_{\tilde{q}_i},m_{q_I}} \right.
\non\\
&& \hspace{1cm} -\  M_{\tilde{\chi_n}}  
 \left( \widetilde{X}^{q_I}_i{G}^{q_I}_i-\eps_n \eps_g {X}^{{q_I}{{}^\dagger}}_i \widetilde{G}^{{q_I}{{}^\dagger}}_i\right)
 \left(C_{23}+C'_{23}+C_{12}\right)\Big|_{M_{\tilde{q}_i},m_{q_I}} 
\non\\
&&\left.\hspace{1cm}\vphantom{\int} -\ m_{q_I}\ 
\left(  X^{q_I}_i{G}^{q_I}_i-\eps_n \eps_g \widetilde{X}^{{q_I}{{}^\dagger}}_i \widetilde{G}^{{q_I}{{}^\dagger}}_i\right)
\  C_0\Big|_{M_{\tilde{q}_i},m_{q_I}} \right).
\label{Cn}
\eea
We have introduced $\eps_n$ and $\eps_g$, which denote the sign of
the mass term of the $n^{th}$ neutralino and of the gluino respectively.
The sum runs through all quarks and squarks.
 
The functions $C_{0/12/23}\Big|_{M_{\tilde{q}_i},m_q}= 
C_{0/12/23}(0,\mchi^2,\mgss^2,m_q^2,m_q^2,\msqi^2)\ $ 
are three-point one-loop Passarino-Veltman functions
\cite{passvelt}\footnote{We follow the conventions and definitions in
  \cite{hlogan}}. The prime denotes the interchange between $m_q^2$ and $\msqi^2$ in the argument.\\

\item $\tilde{g}\to q_{{}_I}\bar{q}_{{}_J} \chi^o_{\rm n} $

\begin{figure}[t]
\centering
\includegraphics[width=15cm]{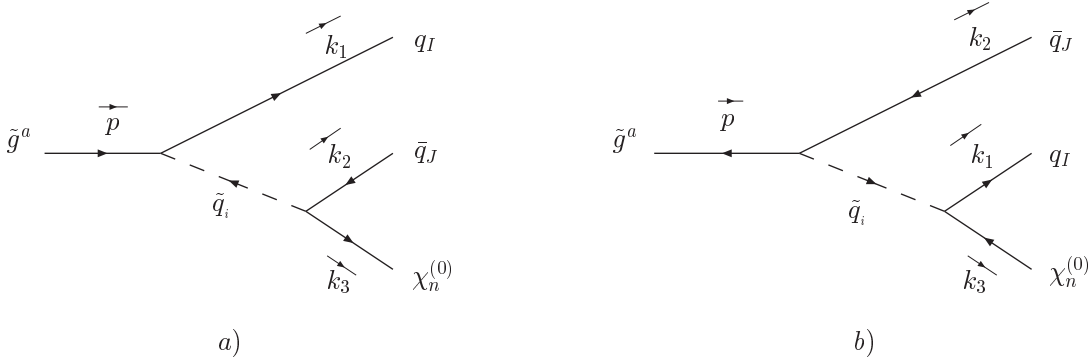}
\caption{Diagrams contributing to the three-body gluino decay to neutralino in MSSM.}
\label{treediags}
\vspace{.3cm}
\end{figure}

The decay width for this process (see Figure~\ref{treediags}) is
{\small\bea
\hspace{-.5cm}\Gamma(\tilde{g}\to\chi_{\rm n}^{o}q_{{}_I} \bar{q}_{{}_J} )
=  {g_s^2\over256 \pi^3 \mgss^3 } \sum_{i,j} \int ds_{13} ds_{23}\ {1\over2}
{\cal R}e\Big(A_{ij}(s_{23}) + B_{ij}(s_{13}) -  2 \eps_{n} \eps_{{g}}\
C_{ij}(s_{23},s_{13})\Big)\ \ \ \ \ \ 
\label{neutralinowidth}
\eea }
where the integrand is the square of the spin-averaged total amplitude
and $i,j=1,2,..,6$ are the indices of the squarks mediating the decay.

The limits of integration are
\eit\vspace{-.2cm}
{\small\bea
\hspace{-.3cm}s_{13}^{max}(s_{23})\!\! &=&\!\! m_{q_{{}_I}}^2 + \mchi^2 +{1\over2
  s_{23}}\left[(\mgss^2-m_{q_{{}_I}}^2-s_{23})(s_{23}-m_{q_{{}_J}}^2+\mchi^2)
  +\lambda^{1/2}(s_{23},\mgss^2,m_{q_{{}_I}}^2)\ \lambda^{1/2}(s_{23},m_{q_{{}_J}}^2,\mchi^2)\right]\non \\
\hspace{-.3cm}s_{13}^{min}(s_{23})\!\! &=&\!\! m_{q_{{}_I}}^2 + \mchi^2 +{1\over2
  s_{23}}\left[(\mgss^2-m_{q_{{}_I}}^2-s_{23})(s_{23}-m_{{q_{{}_J}}}^2+\mchi^2)
  -\lambda^{1/2}(s_{23},\mgss^2,m_{q_{{}_I}}^2)\ \lambda^{1/2}(s_{23},m_{q_{{}_J}}^2,\mchi^2)\right] \non\\
\hspace{-.3cm}s_{23}^{max}\ \ \  &=& (\mgss - m_{q_{{}_I}})^2\non  \\
\hspace{-.3cm}s_{23}^{min}\ \ \  &=& (\mchi + m_{q_{{}_J}})^2 
\eea} where   $\ \lambda(x,y,z)=x^2+y^2+z^2-2 xy-2xz -2yz $ and the
kinematical variables are $s_{13}=(k_1+k_3)^2$ and $s_{23}=(k_2+k_3)^2$.

The $A_{ij}$ terms in Eq.~(\ref{neutralinowidth}) represent the contributions from $diagram\ a)$ in
Figure~\ref{treediags}, whereas the $B_{ij}$ terms come from $diagram\ b)$ of
the same figure.
{\footnotesize
\bea
\hspace{-1.2cm}A_{ij}\!\!\!&=&\!\!\! {\Big( {1\over2}(\mgss^2 +m_{{}_I}^2 - s_{23}) {\rm Tr}\left[{G_i^{q_{{}_I}}G_j^{{q_{{}_I}}{{}^\dagger}}}\right]+m_{{}_I} \mgss 
{\rm Tr}\left[{G_i^{q_{{}_I}}\widetilde{G}_j^{{q_{{}_I}}{{}^\dagger}}}\right] \Big) 
\Big({1\over2}(s_{23}-\mchi^2 -m_{{}_J}^2) {\rm Tr}\left[{X^{q_{{}_J}}_iX_j^{q_{{}_J}{}^\dagger}}\right]-m_{{}_J} \mchi 
{\rm Tr}\left[{X^{q_{{}_J}}_i\widetilde{X}_j^{q_{{}_J}{}^\dagger}}\right] \Big) 
 \over\left(s_{23}-M_{\tilde{{q}}_i}^2\right)\left(
 s_{23}-M_{\tilde{{q}}_j}^2     \right) }\ \ \ \ \ \ \ \ \non\\
\hspace{-1.2cm}B_{ij}\!\!\!&=&\!\!\!  {\Big( {1\over2}(\mgss^2 +m_{{}_J}^2 - s_{13}) {\rm Tr}\left[{G_i^{q_{{}_J}}G_j^{{q_{{}_J}}{{}^\dagger}}}\right]+m_{{}_J} \mgss 
{\rm Tr}\left[{G_i^{q_{{}_J}}\widetilde{G}_j^{{q_{{}_J}}{{}^\dagger}}}\right] \Big) 
\Big({1\over2}(s_{13}-\mchi^2 -m_{{}_I}^2) {\rm Tr}\left[{X^{q_{{}_I}}_iX_j^{q_{{}_I}{}^\dagger}}\right]-m_{{}_I} \mchi 
{\rm Tr}\left[{X^{q_{{}_I}}_i\widetilde{X}_j^{q_{{}_I}{}^\dagger}}\right] \Big) 
 \over\left(s_{13}-M_{\tilde{{q}}_i}^2\right)\left(
 s_{13}-M_{\tilde{{q}}_j}^2     \right) }\non
\eea}
 The $C_{ij}$'s represent the interference terms: 
{\footnotesize
\bea
\hspace{-13.2cm} C_{ij}&=&{T_{ij}  \over  \left(s_{23}-M_{\tilde{{q}}_i}^2\right)
  \left(s_{13}-M_{\tilde{{q}}_j}^2  \right)}
\eea}
with $T_{ij}$ defined by
{\footnotesize
\bea
\hspace{-1.2cm}T_{ij}\!\!\!&=&\!\!\! K_1^{ij} \Big[(s_{13}\! -\!  \mchi^2\! -\! m_{q_{I}}^2)(\mgss^2\! +\!  m_{q_{J}}^2\!  -\! s_{13})\! 
+\!  (s_{23}\!  -\! \mchi^2\! -\! m_{q_{J}}^2)(\mgss^2\! +\! m_{q_{I}}^2 \! -\!  s_{23})\! 
-\! (\mgss^2\! +\! \mchi^2 \!  -\! s_{23}\! -\! s_{13} )(s_{23}\! +\! s_{13}\! -\! m_{q_{I}}^2-\! m_{q_{J}}^2 )\Big]\non\\
\hspace{-1.2cm}\ &&-\  4 \mchi \mgss m_{q_{J}} m_{q_{I}}\ K^{ij}_2 
+\ 2\mgss m_{q_{J}} \left(s_{13}-\mchi^2 -m_{q_{I}}^2\right)\  K^{ij}_3\
+\ 2m_{q_{I}}m_{q_{J}}\left(s_{23}+s_{13}-m_{q_{I}}-m_{q_{J}} \right)\  K_4^{ij}\ \non\\
\hspace{-1.2cm}\ &&+\ 2\mgss m_{q_{I}}\left(s_{23}-\mchi^2-m_{q_{J}}^2 \right)\ K_5^{ij}\
-\ 2\mchi m_{q_{J}}\left(\mgss^2+m_{q_{I}}^2 -s_{23}\right)\ K_6^{ij}\ \non\\
\hspace{-1.2cm}\ && -\ 2\mchi\mgss \left(\mgss^2+\mchi^2 -s_{13}-s_{23}\right)\ K_7^{ij}\
-\ 2\mchi m_{q_{I}} \left(\mgss^2+m_{q_{J}}^2-s_{13}\right)\ K_8^{ij}\ 
\eea}
The eight types of effective coupling constants $K_a^{ij}$ ($a=1,..,8$) appearing in these
terms depend on the original ouplings as
{\small\bea
&& K_1^{ij}={1\over4}{\rm Tr}\left[X^{q_{{}_J}}_i\widetilde{X}_j^{q_{{}_I}{}^\dagger}  G_i^{q_{{}_I}}\widetilde{G}_j^{q_{{}_J}{{}^\dagger}} \right]  
{\rm ,} \hspace{.5cm} 
K_2^{ij}=  {1\over4}{\rm Tr}\left[X^{q_{{}_J}}_i X_j^{q_{{}_I}{}^\dagger} G_i^{q_{{}_I}} G_j^{q_{{}_J}{{}^\dagger}} \right] 
{\rm ,} \hspace{.5cm} 
K_3^{ij}= {1\over4} {\rm  Tr}\left[X^{q_{{}_J}}_i \widetilde{X}_j^{q_{{}_I}{}^\dagger}{G}_i^{q_{{}_I}} {G}_j^{q_{{}_J}{{}^\dagger}}  \right] \non\\
&& K_4^{ij}=  {1\over4} {\rm  Tr}\left[X^{q_{{}_J}}_i\widetilde{X}_j^{q_{{}_I}{}^\dagger}\widetilde{G}_i^{q_{{}_I}} G_j^{q_{{}_J}{{}^\dagger}} \right]
 {\rm , } \hspace{.5cm} 
K_5^{ij}= {1\over4} {\rm  Tr}\left[X^{q_{{}_J}}_i \widetilde{X}_j^{q_{{}_I}{}^\dagger} \widetilde{G}_i^{q_{{}_I}} \widetilde{G}_j^{q_{{}_J}{{}^\dagger}} \right]
{\rm ,} \hspace{.5cm} 
K_6^{ij}= {1\over4} {\rm  Tr}\left[ X^{q_{{}_J}}_i X_j^{q_{{}_I}{}^\dagger} \widetilde{G}_i^{q_{{}_I}} G_j^{q_{{}_J}{{}^\dagger}}  \right]  \non\\
&& K_7^{ij}=  {1\over4} {\rm  Tr}\left[ X^{q_{{}_J}}_i X_j^{q_{{}_I}{}^\dagger}  \widetilde{G}_i^{q_{{}_I}}  \widetilde{G}_j^{q_{{}_J}{{}^\dagger}}  \right]
\hspace{.5cm} {\rm and} \hspace{.5cm} 
K_8^{ij}= {1\over4} {\rm  Tr}\left[X^{q_{{}_J}}_i X_j^{q_{{}_I}{}^\dagger}G_i^{q_{{}_I}}\widetilde{G}_j^{q_{{}_J}{{}^\dagger}}  \right]\non
\eea}
where the traces are computed in Dirac Space.

\bit
\item $\tilde{g}\to d_I \bar{u}_J \chi^+_{\rm m}$

\begin{figure}[t]
\centering
\includegraphics[width=15cm]{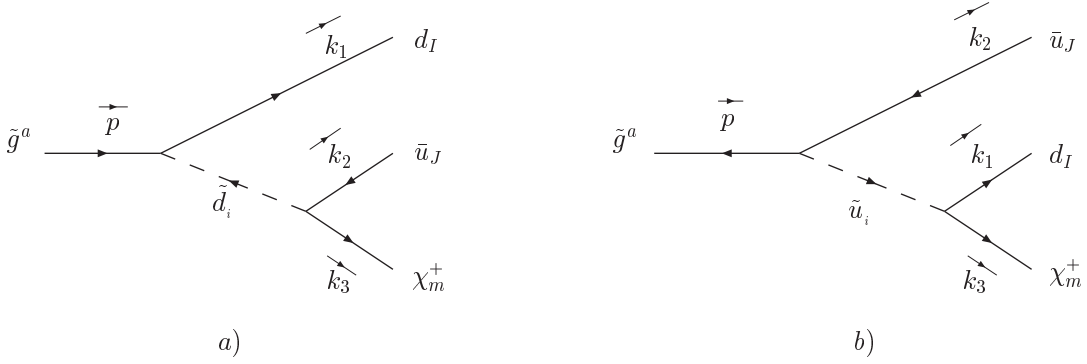}
\caption{Diagrams contributing to the gluino decay to chargino in the MSSM.}
\label{chargitreediags}
\vspace{.3cm}
\end{figure}

The decay width of this process (see Figure~\ref{chargitreediags}) is
\bea
\Gamma(\tilde{g}\to\chi_{\rm m}^{+}d_{{}_I}\bar{u}_{{}_J}  )
=  {g_s^2\over256 \pi^3 \mgss^3 } \sum_{i,j} \int ds_{13} ds_{23}\ {1\over2}
{\cal R}e\Big(A'_{ij}(s_{23}) + B'_{ij}(s_{13}) -  2  \eps_{\tilde{g}}\
C'_{ij}(s_{23},s_{13})\Big)\ \ \ \ \ \ \ \
\label{charginowidth}
\eea 
where the integrand is the square of the spin-averaged total amplitude
and $i,j=1,2,..,6$ are the indices of the squarks mediating the decay.

The limits of integration are
\eit
{\small
\bea
s_{13}^{max}(s_{23})\!\!\! &=&\!\!\! m_{d_{{}_I}}^2\! + \mchi^2 +{1\over2
  s_{23}}\left[(\mgss^2-m_{d_{{}_I}}^2\!-s_{23})(s_{23}-m_{u_{{}_J}}^2\!+\mchi^2)
  +\lambda^{1/2}(s_{23},\mgss^2,m_{d_{{}_I}}^2)\ \lambda^{1/2}(s_{23},m_{u_{{}_J}}^2,\mchi^2)\right]\non \\
s_{13}^{min}(s_{23})\!\!\! &=&\!\!\! m_{d_{{}_I}}^2\! + \mchi^2 +{1\over2
  s_{23}}\left[(\mgss^2-m_{d_{{}_I}}^2\!-s_{23})(s_{23}-m_{u_{{}_J}}^2\!+\mchi^2)
  -\lambda^{1/2}(s_{23},\mgss^2,m_{d_{{}_I}}^2)\ \lambda^{1/2}(s_{23},m_{u_{{}_J}}^2,\mchi^2)\right] \non\\
s_{23}^{max}\ \ \  &=& (\mgss - m_{d_{{}_I}})^2\non  \\
s_{23}^{min}\ \ \  &=& (\mchi + m_{u_{{}_J}})^2 ,
\eea}
where   $\ \lambda(x,y,z)=x^2+y^2+z^2-2 xy-2xz -2yz $ and the
kinematical variables are $s_{13}=(k_1+k_3)^2$ and $s_{23}=(k_2+k_3)^2$.

The terms ${A}'_{ij}$, ${B}'_{ij}$ and ${C}'_{ij}$ in
Eq.~(\ref{charginowidth}) are 
{\footnotesize
\bea
\hspace{-1.4cm}{ A}'_{ij}\!\!\!\!&=&\!\!\!\! {\Big( {1\over2}(\mgss^2 +m_{d_{{}_I}}^2\! - s_{23}) {\rm Tr}\left[{G_i^{d_{{}_I}}G_j^{{d_{{}_I}}{{}^\dagger}}}\right]\!\!+m_{d_{{}_I}}\! \mgss 
{\rm Tr}\left[{G^{d_{{}_I}}_i\widetilde{G}_j^{{d_{{}_I}}{{}^\dagger}}}\right] \Big) 
\Big({1\over2}(s_{23}-\mchi^2 -m_{u_{{}_J}}^2) {\rm Tr}\left[{Y^{u_{{}_J}}_iY_j^{{u_{{}_J}}{{}^\dagger}}}\right]\!\!-m_{u_{{}_J}}\! \mchi 
{\rm Tr}\left[{Y^{u_{{}_J}}_i\widetilde{Y}_j^{{u_{{}_J}}{{}^\dagger}}}\right] \Big) 
 \over\left(s_{23}-M_{\tilde{d}_i}^2\right)\left(
 s_{23}-M_{\tilde{d}_j}^2     \right) }\non\\
\hspace{-1.4cm}{ B}'_{ij}\!\!\!\!&=&\!\!\!\! {\Big( {1\over2}(\mgss^2 +m_{u_{{}_J}}^2\! - s_{13}) {\rm Tr}\left[{G^{u_{{}_J}}_iG_j^{{u_{{}_J}}{{}^\dagger}}}\right]+\!\!m_{u_{{}_J}}\! \mgss 
{\rm Tr}\left[{G^{u_{{}_J}}_i\widetilde{G}_j^{{u_{{}_J}}{{}^\dagger}}}\right] \Big) 
\Big({1\over2}(s_{13}-\mchi^2 -m_{d_{{}_I}}^2) {\rm Tr}\left[{Y_i^{d_{{}_I}}Y_j^{{d_{{}_I}}{{}^\dagger}}}\right]-\!\!m_{d_{{}_I}}\! \mchi 
{\rm Tr}\left[{Y_i^{d_{{}_I}}\widetilde{Y}_j^{{d_{{}_I}}{{}^\dagger}}}\right] \Big) 
 \over\left(s_{13}-M_{\tilde{t}_i}^2\right)\left(
 s_{13}-M_{\tilde{t}_j}^2     \right) }\non
\eea}
and
{\footnotesize
\bea
\hspace{-13cm}C'_{ij}& =&{T'_{ij} \over  \left(s_{23}-M_{\tilde{b}_i}^2\right)
  \left(s_{13}-M_{\tilde{t}_j}^2  \right)}
\eea}
with
{\footnotesize
\bea
\hspace{-1.5cm} T'_{ij}\!\!\!\!&=&\!\!\!\!{K'_1}^{ij}   \Big[(s_{13}\! -\! \mchi^2\! -\!  m_{d_{{}_I}}^2)(\mgss^2\!+ \! m_{u_{{}_J}}^2\!\!  -\! s_{13})\!+\! (s_{23}\!  -\!
  \mchi^2\! -\! m_{u_{{}_J}}^2)(\mgss^2\! +\! m_{d_{{}_I}}^2\!\! -\! s_{23})\!  -\!  (\mgss^2\! +\! \mchi^2\!  -\! s_{23}\! -\! s_{13} )(s_{23}\! +\!
  s_{13}\! -\! m_{d_{{}_I}}^2\!\! -\! m_{u_{{}_J}}^2 )\Big]\ \non\\
\hspace{-1cm}\ && -\  4\mchi \mgss m_{d_{{}_I}} m_{u_{{}_J}}\ {K'_2}^{ij}\! 
+\ 2\mgss m_{u_{{}_J}}  \left(s_{13}-\mchi^2 -m_{d_{{}_I}}^2\right)\ {K'_3}^{ij}
+\ 2m_{d_{{}_I}} m_{u_{{}_J}}\left(s_{23}+s_{13}-m_{d_{{}_I}}^2-m_{u_{{}_J}}^2\right)\ {K'_4}^{ij} \non\\
\hspace{-1cm}\ &&+\ 2\mgss m_{d_{{}_I}}\left(s_{23}-\mchi^2-m_{u_{{}_J}}^2 \right)\ {K'_5}^{ij}
 -\ 2\mchi m_{u_{{}_J}}  \left(\mgss^2+m_{d_{{}_I}}^2 - s_{23}\right)\ {K'_6}^{ij}\non\\
\hspace{-1cm}\ &&-\ 2\mchi \mgss \left(\mgss^2+\mchi^2-s_{13}-s_{23}\right) \ {K'_7}^{ij}
-\ 2\mchi m_{d_{{}_I}} \left(\mgss^2+m_{u_{{}_J}}^2-s_{13}\right)\ {K'_8}^{ij} 
\eea}
The effective couplings ${K'_a}^{ij}$ are
\bea
&& {K'_1}^{ij}={1\over4}{\rm Tr}\left[Y_i^{u_{{}_J}}Y_j^{{d_{{}_I}}} G_i^{d_{{}_I}}\widetilde{G}_j^{{u_{{}_J}}{{}^\dagger}} \right] 
{\rm ,} \hspace{.5cm} 
{K'_2}^{ij}=  {1\over4}{\rm Tr}\left[Y_i^{u_{{}_J}} \widetilde{Y}_j^{{d_{{}_I}}} G_i^{d_{{}_I}} G_j^{{u_{{}_J}}{{}^\dagger}} \right]
{\rm ,} \hspace{.5cm} 
{K'_3}^{ij}= {1\over4}     {\rm  Tr}\left[Y_i^{u_{{}_J}} Y_j^{{d_{{}_I}}}{G}_i^{d_{{}_I}} {G}_j^{{u_{{}_J}}{{}^\dagger}}  \right]      \non\\
&& {K'_4}^{ij}=  {1\over4}  {\rm  Tr}\left[Y_i^{u_{{}_J}}Y_j^{{d_{{}_I}}}\widetilde{G}_i^{d_{{}_I}} G_j^{{u_{{}_J}}{{}^\dagger}} \right]
 {\rm , } \hspace{.5cm} 
{K'_5}^{ij}= {1\over4}  {\rm  Tr}\left[Y_i^{u_{{}_J}} Y_j^{{d_{{}_I}}} \widetilde{G}_i^{d_{{}_I}} \widetilde{G}_j^{{u_{{}_J}}{{}^\dagger}} \right]
{\rm ,} \hspace{.5cm} 
{K'_6}^{ij}= {1\over4}  {\rm  Tr}\left[ Y_i^{u_{{}_J}} \widetilde{Y}_j^{{d_{{}_I}}} \widetilde{G}_i^{d_{{}_I}} G_j^{{u_{{}_J}}{{}^\dagger}}  \right]  \non\\
&& {K'_7}^{ij}=  {1\over4}  {\rm  Tr}\left[ Y_i^{u_{{}_J}} \widetilde{Y}_j^{{d_{{}_I}}}  \widetilde{G}_i^{d_{{}_I}}  \widetilde{G}_j^{{u_{{}_J}}{{}^\dagger}}  \right]
\hspace{.5cm} {\rm and} \hspace{.5cm} 
{K'_8}^{ij}= {1\over4}  {\rm  Tr}\left[Y_i^{u_{{}_J}} \widetilde{Y}_j^{{d_{{}_I}}}G_i^{d_{{}_I}}\widetilde{G}_j^{{u_{{}_J}}{{}^\dagger}}  \right]
\eea
where the traces are computed in Dirac Space.

\section{Application to Split Supersymmetry} \label{sectionpheno}

We wish now to compute the branching fractions of the gluino in
some examples of split supersymmetry.  Our intention here is to not
delve into various model building aspects of split supersymmetry,
but to give the reader an understanding for how very different the possibilities 
can be for gluino decay phenomenology if the scalars are much heavier
than the gauginos.

Despite the fact that we are interested in general low-scale parameter
descriptions of the phenomenology of split supersymmetry, it is helpful
at times to give names to various orderings of the gaugino mass parameters.
For example, in ``minimal supergravity'' (mSUGRA) and in ``anomaly mediated
supersymmetry breaking'' (AMSB), there are particular
orderings of the gaugino masses:
\bea
{\rm mSUGRA}~ \left\{ \begin{array}{l} 
M_1\ =\ 0.4\ M_{1/2} \\       
M_2\ =\ 0.8\ M_{1/2}   \\     
M_3\ =\ \ 3\ M_{1/2}\end{array}\right. \hspace{1cm}{\rm and} \hspace{1cm}
{\rm AMSB}~\left\{ \begin{array}{l} 
M_1\ =\ 3 M_{2} \\       
M_2\ =\ \ M_{2}   \\     
M_3\ =\ 7 M_{2}\end{array}\right.             
\eea

For simplicity of our illustrations,
we also define a common scalar mass $m_0$ which corresponds to the
mass of all squarks with the exception of $\tilde{t}_R$, $\tilde{b}_R$ 
and $\tilde{Q}^{3}_L$.  
Inspired by the usual RGE effects on scalar masses, we will take the
third family squarks slightly lower than $m_0$ depending on the value of
$\tan{\beta}$.
More specifically we will take\footnote{These numbers are roughly expected
if we do a renormalization group flow of common scalar masses
$m_0=10^4$-$10^5$ GeV from the 
GUT scale down to the weak scale.}
\bea
\begin{array}{c|ccc}
\tan{\beta}&m_{\tilde{Q}^3_L}&m_{\tilde{b}_R} &m_{\tilde{t}_R}\\\hline\\
30\ & 0.8\ m_0\ &\ 0.9\ m_0\ &\ 0.6\ m_0\\
1.5\ & 0.7\ m_0\ &\ 1\  m_0\ &\ 0.4\ m_0
\end{array}
\eea
The trilinear parameters are ignored in the scan and nominally set to
$200$ GeV when a precise value is needed.
With these input parameters, all mixings and physical masses are determined and
inserted in the formulae given in the first section of the present paper.

\subsection{Importance of 2-body decays}

\begin{figure}[t]
\centering
\includegraphics[width=5.5cm]{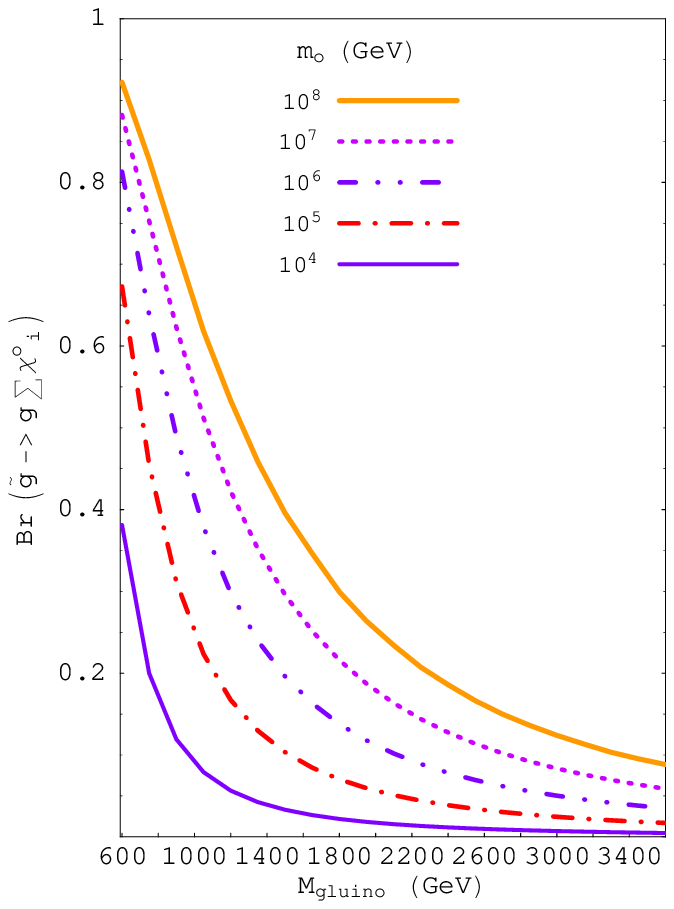}
\includegraphics[width=5.5cm]{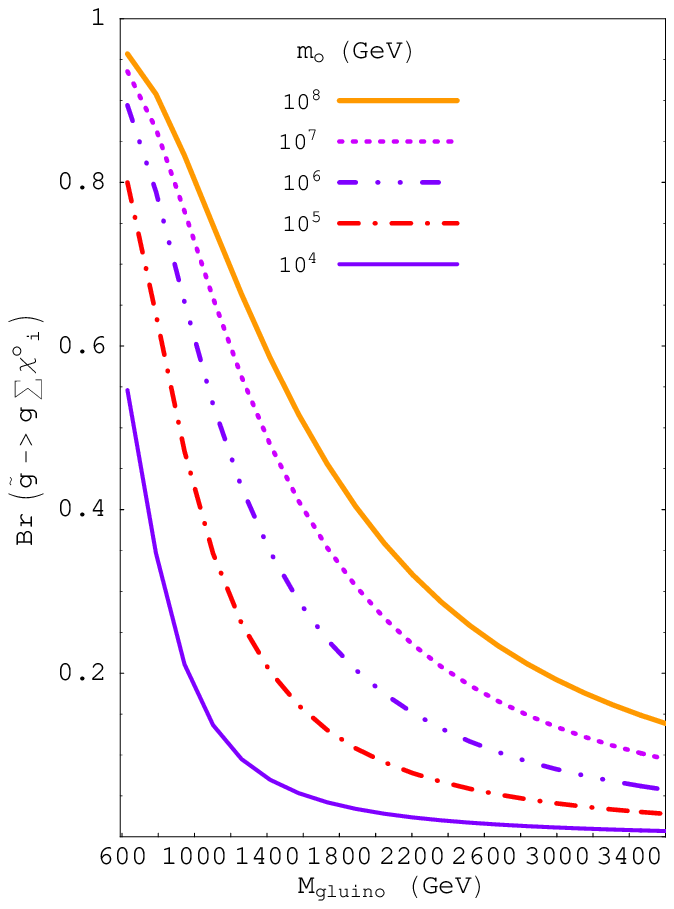}
\caption{Radiative two-body branchings of the gluino in
  mSUGRA (left) and AMSB (right) with $\mu=M_1$ and $\tan\beta=1.5$.
These two plots illustrate the argument in the text that the heavier the
$gluino$ mass, the larger the scalar masses need to be for the two-body decay
to be sizeable.}
 \label{BR2Bplots}
\end{figure}

The ratio $R_{2/3}={\Gamma(\tilde{g}\to {\rm 2\ Body}) \over
  \Gamma(\tilde{g}\to {\rm 3\ Body})}$ between the radiative 2-body 
decay width of the
gluino and its total 3-body decay width into neutralinos and charginos
starts to have a non-trivial scaling when the squarks mediating the
decays become very heavy. When the gluino is kinematically allowed to
  decay into Higgsinos, $R_{2/3}$ scales with the gluino and
  scalar masses as
\bea
R_{2/3} \propto 
{ 
m_t^2 \left(1-\log{{M^2_{\tilde{t}}\over m_t^2}}\right)^2\over \mgss^2 }
\eea
The large $Log$ from the radiative 2-Body decay width appears out of the
Passarino-Veltman function $C_0$ in Eq.~(\ref{Cn}), in the limit of large
squark mass.

This $Log$ enhancement comes from the Higgsino coupling to the internal
quark and squark running in the loop. It can be understood from
the effective theory point of view after integrating out the
heavy scalars.  In that case, a four-point fermion interaction
of quark-quark-gluino-higgsino can have its two quark lines tied
together and a gluon can radiate from any strongly interacting particle.
This diagram is divergent in the effective theory, which is cut off
by the squark mass (the scale of the effective theory breakdown).
The analagous construction of the two body decay from
the quark-quark-gluino-wino diagram in the effective theory has
no divergence, and therefore no log of the squark mass.

Thinking of these decays within the effective theory description
enables us to understand that the purely diagrammatic calculation
presented in this paper (and the simpler equivalent calculations
in previous papers) are not entirely adequate when the scalar masses
are much heavier than the gaugino masses.  The large logarithm can cause
a breakdown in perturbation theory for the diagrammatic calculation.
To do the calculation properly in that case would require matching
the effective theory with the full theory at the heavy scalar mass
scale and performing a renormalization group running of operator
coefficients down to the gluino scale and then computing the decay
in the effective theory.  We have estimated that as long as
$m_0<10^8\gev$ (for $M_{gluino}$ less than a few TeV)
we do not have to worry about this effect, and that is one reason
we are cutting off all our graphs at that scale.  This does not
cause us much concern in our analysis as it is our opinion that the
most straightforward split supersymmetry scenarios have scalar masses
only a loop level (or so) higher than the gaugino masses, which is
well within the confines of our graphs.

Therefore, when the squark masses are sufficiently large, the 
logarithm can
overcome any loop factor supression and the supression from the
gluino mass squared in the denominator. Of course the heavier the
gluino, the larger the scalar masses need to be for the two-body decay
to be sizeable. This situation is
well represented in Figure~\ref{BR2Bplots}, where the 2-body
loop-induced branching fraction of gluino decay becomes smaller
as the gluino mass increases for fixed scalar mass $m_0$. 
We take the two different low scale spectra defined
earlier, mSUGRA with $M_{1/2}=300$ GeV and AMSB with $M_2=120$
GeV. In both cases we take for this plot $\tan{\beta}=1.5$ and 
$\mu=M_1$. 

Note, the $Log$ dominance occurs when the higgsino is kinematically
allowed in gluino decays.  If that is not the case, the lack of a
$Log$ enhancement of the two-body decays means the three-body decay
will generally win out.  Therefore, the mass of the higgsino is of
prime importance to gluino decay phenomenology.

\subsection{Gluino decay phenomenology}

We proceed now to describe the behavior of different branching ratios of
gluino decay when the scalar mass parameter $m_0$ is varied between the TeV
scale to $10^8$ GeV. In all plots a gray vertical band is located
roughly between $m_0=10^6$ GeV to $m_0=10^7$ GeV. This band corresponds
to the range of the total gluino decay width such that the $c\tau$ (in
rest frame) is between $1mm$ and $10 m$, and therefore
roughly gives an estimate of where displaced vertices can be seen. To
the right of the band the gluino is effectively stable (as far as the
detector is concerned) and the phenomenology of that situation is
studied in~\cite{stable gluino}. To the left of the band the gluino decays promptly,
but the decay modes can change dramatically, depending on the scenario
and the value of $m_0$, mostly due to the emergence of the
2-body decay channel into gluon and Higgsino.\\

\begin{figure}[t]
\vspace{-0cm}
\centering
\includegraphics[width=5.5cm]{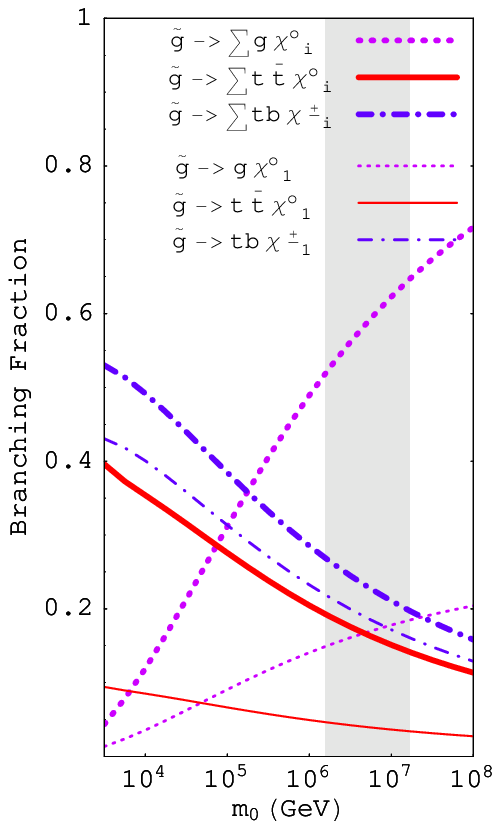}
\includegraphics[width=5.5cm]{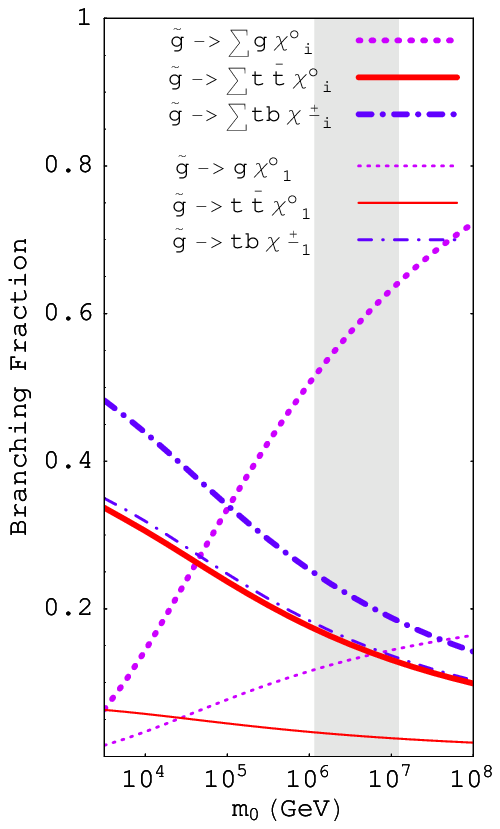}
\vspace{-.5cm}
\caption{Branching Fractions of gluino decay for mSUGRA with
  $m_{1/2}=300$ GeV and $\mu=120$ GeV and with $\tan\beta=1.5$ (left)
  and $\tan\beta=30$ (right).  The heavy scalar mass $m_0$ enables
larger two-body final state branching fractions, as the Higgsino+gluon
final state is enhanced by $\log m_0$ over other decay channels.  The shaded
band in the figure represents $1\,{\rm mm}<c\tau_{\tilde g} < 10\,{\rm m}$.}
\label{sugra120}
\end{figure}

The input parameters of Figure~\ref{sugra120}, come from mSUGRA-like 
low-scale relations
between gaugino mass parameters, with $M_{1/2}=300$ GeV and $\mu=120$
GeV.
In the left figure $\tan\beta=1.5$ which also means that we
take the stop mass to be $0.4 m_0$, slightly enhancing the stop mediated
decays. 
In the right figure we take $\tan\beta=30$ to compare the enhancement
effect. 
Another feature of the spectrum taken for these plots is the value of
$\mu=120$ GeV such that the lightest neutralino is a mixed state of 
Bino and Higgsino.

In the figures we plot two types of lines, thick and thin.
The thick dotted line gives the Branching fraction of gluino decay into
gluon plus (any) neutralino, the thick solid line represents the decay
into $t \bar{t}$ plus (any) neutralino and the thick dash-dotted line
gives the $top$-$bottom$ plus (any) chargino channel.

The thin lines are really $exclusive$ and concentrate on the lightest
neutralino and chargino, so that the thin dotted line represents decay
into gluon plus lightest neutralino ($LSP$), the thin solid 
represents $t\bar{t}$
plus lightest neutralino and the thin dash-dotted line represents
$bottom$-$top$ plus lightest chargino.

The two plots of Figure~\ref{sugra120} show clearly that in the PeV 
scale range the 2-body
decay starts to dominate over chargino and neutralino 3-body decays. Not
surprisingly the decay into $t\bar{t}$ and $bt$ is enhanced for small
$\tan\beta$ since the stop mediating that decay channel is clearly lighter than
the rest of squarks. The $t\bar{t}$ plus neutralino Branching ranges 
from $0.4$ to $0.2$
in the prompt decay zone of the gluino, and can thus be an interesting signal,
since between 15\% and 5\% of the time a pair of gluinos will give rise to at
least 4 top events.

As for the exclusive signals we see that the $t\bar{t}$ plus $LSP$
branching ranges between $0.1$ and $0.05$ and monotonically decreases with
$m_0$ due to the fact that the 2-body branching is increasing.
The other interesting signal is the 2-body decay into gluon and $LSP$,
the branching of which increases to values of around $0.13-0.17$ for the
larger scalar masses. This mode has some importance in this case because
the lightest neutralino carries some Higgsino component and therefore
its radiative gluino decay increases thanks to the $Log$ enhacement
discussed in the previous section.
A pair of gluinos decaying into two jets plus substantial
$E\hspace{-.27cm}/$ (from two $LSP$'s) would be an interesting and
unexpected result from supersymmetry.

\begin{figure}[t]
\centering
\includegraphics[width=5.5cm]{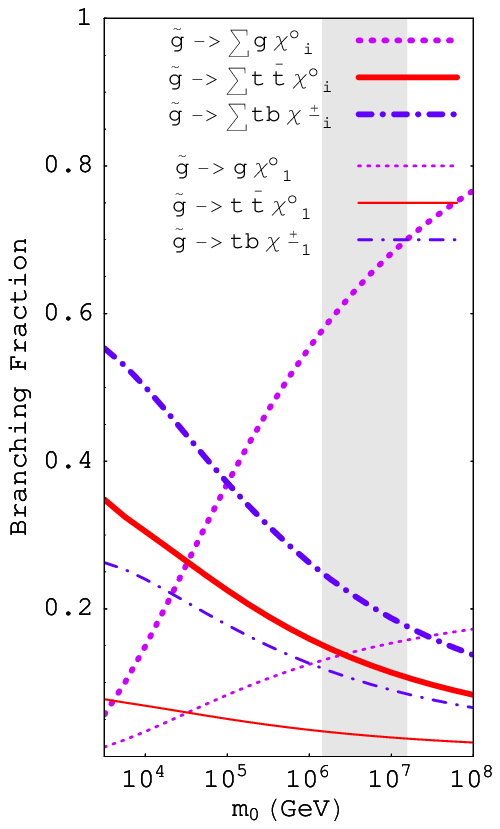}
\includegraphics[width=5.5cm]{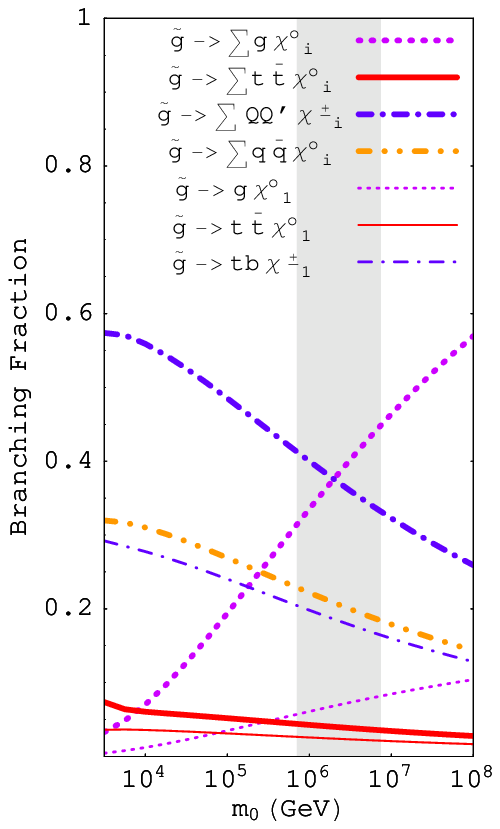}
\vspace{-.5cm}
\caption{$M_{2}=120 $, $ \tan\beta=1.5$   $ \mu= 120$ (left),  
$ \mu= 700$ (right)  AMSB.  The heavy scalar mass $m_0$ enables
larger two-body final state branching fractions, as the Higgsino+gluon
final state is enhanced by $\log m_0$ over other decay channels.  The shaded
band in the figure represents $1\,{\rm mm}<c\tau_{\tilde g} < 10\,{\rm m}$.
The two panels of this figure differ in $\mu$.  As $\mu$ increase (right panel)
the ability to decay into Higgsinos is diminished and the two-body final
state branching fraction is reduced. In the right panel, lower case $q$
means that only the five lighter quarks, $u,d,c,s,b$ are included in the decay
channel, while upper case $Q$ means all six Standard Model quarks
are included.}
\label{amsb generic}
\end{figure}

In Fig.~\ref{amsb generic} we show the generic case for gluino decays
when the gauginos follow an AMSB mass ordering.  Again, there are many
final states to untangle at the accelerator to determine exactly
how the gluino is decaying.  Also, we point out that the two body decay
is decreased in the right panel of this diagram compared to the left
panel due to the higgsino mass being nearly as heavy as the gluino mass.
For higgsino masses above the gluino mass the dotted two body decay
line would drop significantly below the three body decays to charginos.

\begin{figure}[h]
\centering
\includegraphics[width=8cm]{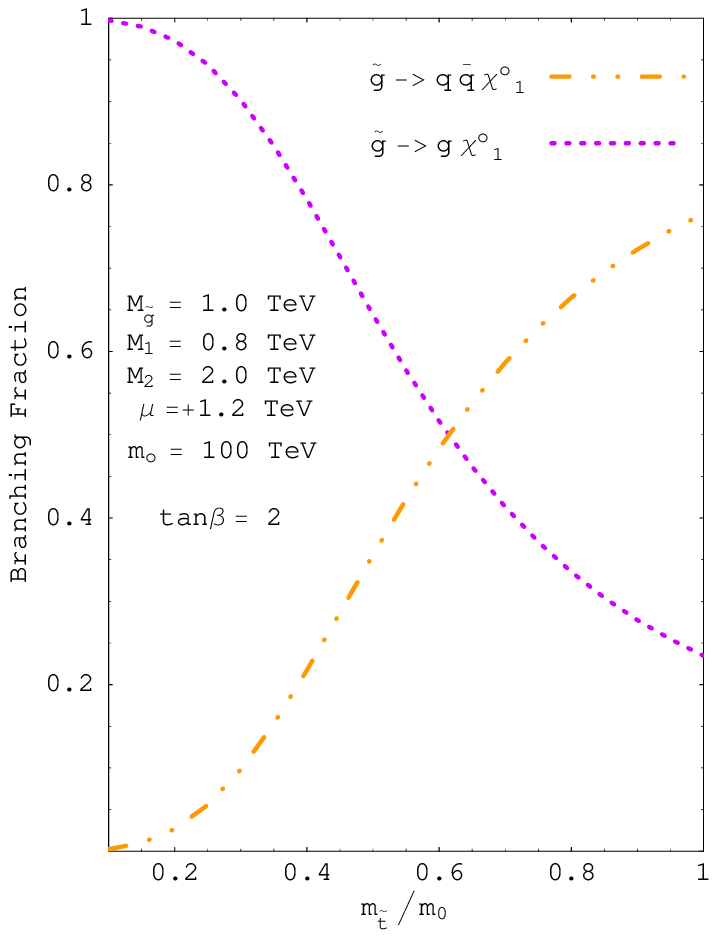}
\includegraphics[width=8cm]{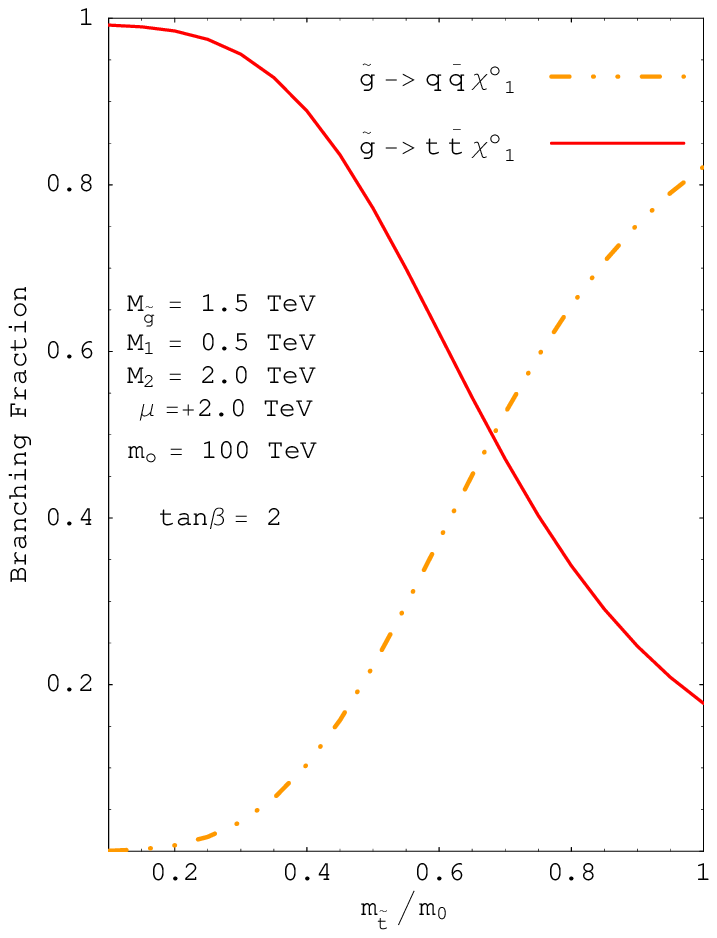}
\caption{The two panels of this figure demonstrate the possibility of 
near 100\% branching fraction of the gluino into a single gluon jet
plus missing energy
($m_{\tilde t}/m_0 <0.2$ in left panel) or near 100\%
branching fraction of the gluino into two top-quarks plus missing energy
($m_{\tilde t}/m_0 <0.2$ in right panel). Lower case $q$ means that only
the five lighter quarks $u,d,c,s,b$ are included in the decay channel}
\label{unique cases}
\end{figure}

Finally, we would like to point out two cases with reasonable parameters
that generate unique final state phenomenology for gluino production
and decay.  The two cases are represented by the two panels of 
Fig.~\ref{unique cases}.  In both of these cases we have chosen
gaugino mass parameters and higgsino mass parameters judiciously,
but not wantonly, to demonstrate that the branching fraction of
the gluino could go into a single final state of considerable challenge
for the LHC.

In the first panel of Fig.~\ref{unique cases} we have a case where
the gluino wants to decay 100\% of the time to gluon plus neutralino
when the top squark mass is about a factor of 5 or more below the
general scalar masses.  Recall this is a reasonable assumption given
renormalization group flow of top squark masses which want to be
driven to lighter values from large top Yukawa coupling. The loop
has light top squarks contributing to them, but the three-body decays
cannot take advantage of the light top squarks since top quarks are
not kinematically allowed in the final state.
Therefore, loop decay
to gluino plus neutralino wins.(When the top squark mass is
larger and comparable to other squark masses, the three body decay
wins out because there is no large relative advantage of the two-body
decay over three-body decays.)  The LHC phenomenology in this case would be
\bea
\tilde g\tilde g\to gg\chi^0_1\chi^0_1~{\rm (two~jets~plus~missing~energy)}
\eea
Determining that this is supersymmetry would be quite a challenge for
experiment.

The second panel of Fig.~\ref{unique cases} is similar to the first
panel except the mass hierarchies are shuffled a bit.  These rather
small changes lead to a huge impact in the final state of gluino decays.
In this case, the top quarks are kinematically allowed in the final state,
and they win dramatically when the top squark masses are somewhat less
than the other squark masses.  The supersymmetry signal of this
theory at the LHC is simply
\bea
\tilde g\tilde g\to t\bar tt\bar t\chi^0_1\chi^0_1
~{\rm (four~top~plus~missing~energy)}
\eea
Determining that there are four top quarks in a final state would be challenging
enough as it is, but to determine there is some missing energy in the
event would increase the difficulty of the experiment.  Given the entire spectrum
of this supersymmetric model, it is possible that four tops plus missing
energy could be the only signal for supersymmetry at the LHC.  As there
are other exotic ideas for producing four top quarks at the LHC, 
establishing that supersymmetry is what we are seeing
would require good ideas and great experimental diligence.

\addcontentsline{toc}{section}{Appendix:}

\section*{Appendix: Couplings used in the numerical analysis  }
\label{appendixA} \renewcommand{\thesubsection}{A.\arabic{subsection}}
\label{appendixA} \renewcommand{\theequation}{A.\arabic{equation}}
\setcounter{equation}{0}
\setcounter{footnote}{1}
\setcounter{subsection}{0}





\bit

\item {\it Gluino couplings $G^{q_{{}_I}}_i$ }

The gluino couplings are 
\bea
G^{q_{{}_I}}_{1}&=& \cos{\theta_q}\ P_R\ - \eps_g\ \sin{\theta_q}\  P_L\\
G^{q_{{}_I}}_2&=&\hspace{-.3cm}-\sin{\theta_q}\ P_R\ - \eps_g\ \cos{\theta_q}\  P_L 
\eea
where $P_{L/R}= {1\over 2}(1\mp \gamma_5)$.

\item {\it Neutralino couplings $X^{q_{{}_I}}_i$ }

The neutralino-quark-squark interactions are
\bea
X^{q_{{}_I}}_1 &=& \cos{\theta_q}\ X^{q_{{}_I}}_{\tilde{q}_L}+ \sin{\theta_q}\ X^{q_{{}_I}}_{\tilde{q}_R}\\
X^{q_{{}_I}}_2 &=&\hspace{-.3cm}- \sin{\theta_q}\ X^{q_{{}_I}}_{\tilde{q}_L}+ \cos{\theta_q}\ X^{q_{{}_I}}_{\tilde{q}_R}
\eea
where we have \cite{Haber-Wyler}
\bea
X^t_{\tilde{t}_L}&=& -{g\over\sqrt{2}} \left[\eps_n  \Big(Z^*_{n2}+
    {1\over3}\tan{\theta_W}\ Z^*_{n1}\Big)\ P_L  +\   {m_t\over m_W \sin{\beta}}
    Z_{n4}\ P_R            \right]    \\
X^t_{\tilde{t}_R}&=& -{g\over\sqrt{2}}\left[ \eps_n   {m_t\over m_W
    \sin{\beta}} Z^*_{n4}\ P_L   -   \Big({4\over3} \tan{\theta_W}\Big)\
  Z_{n1} \ P_R\right]
\eea
for the up-type quark interactions and
\bea
X^b_{\tilde{b}_L}&=& -{g\over\sqrt{2}}\left[\eps_n
\Big(-Z^*_{n2}+{1\over3}\tan{\theta_W}\ Z^*_{n1}\Big)\ P_L\ +\     {m_b\over m_W \cos{\beta}} Z_{n3} \ P_R \right]                 \\
X^b_{\tilde{b}_R}&=& -{g\over\sqrt{2}}\left[   \eps_n    {m_b\over m_W
    \cos{\beta}} Z^*_{n3}\ P_L \      +\ \Big({2\over3}
  \tan{\theta_W}\Big) Z_{n1}\ P_R \right] 
\eea
for the down-type quark.

As usual $P_{L/R}= {1\over 2}(1\mp \gamma_5)$, and the matrix
$Z$ is responsible for turning the neutral higgsinos, bino and
wino into the physical neutralinos. The signs of the resulting
neutralino mass terms are given by $\eps_n$ with $n=1,2,3,4$.

\item {\it Chargino couplings $Y^t_i$ and $Y^b_i$ }

The Chargino-quark-squark interactions become
\bea
Y^{t}_1 &=& \cos{\theta_b}\ Y^t_{\tilde{b}_L}+ \sin{\theta_b}\ Y^t_{\tilde{b}_R}\\
Y^{t}_2 &=&\hspace{-.3cm}- \sin{\theta_b}\ Y^t_{\tilde{b}_L}+ \cos{\theta_b}\ Y^t_{\tilde{b}_R}
\eea
\bea
Y^{b}_1 &=& \cos{\theta_t}\ Y^b_{\tilde{t}_L}+ \sin{\theta_t}\ Y^b_{\tilde{t}_R}\\
Y^{b}_2 &=&\hspace{-.3cm}- \sin{\theta_t}\ Y^b_{\tilde{t}_L}+ \cos{\theta_t}\ Y^b_{\tilde{t}_R}
\eea
And we have \cite{Haber-Gunion} for up-type quarks:
\bea
Y^t_{\tilde{b}_L}&=& -{g\over \sqrt{2}}\ \left[ \sqrt{2}\ U^*_{m1}\ P_L \ -\ {m_t V_{m2}\over
   m_W \sin{\beta}}\eta_m\ P_R\right]\\
Y^t_{\tilde{b}_R}&=& {g\over \sqrt{2}}\  {m_b U^*_{m2}\over  m_W \cos{\beta}}\ P_L
\eea
and for down-type quarks:
\bea
Y^b_{\tilde{t}_L}&=& {g\over \sqrt{2}}\ \left[ {m_b U^*_{m2}\over
   m_W \cos{\beta}}\ P_L\ -\ \sqrt{2}\ V_{m1}\eta_m\  P_R\right] \\
Y^b_{\tilde{t}_R}&=& {g\over \sqrt{2}}\  {m_t V_{m2}\over  m_W
  \sin{\beta}}\eta_m\ P_R  
\eea

The matrices $U$ and $V$ are responsible for diagonalizing the Chargino
mass matrix, and $\eta_m$ with $m=1,2 $, are the signs of the resulting
chargino masses. 
\eit

\section*{Acknowledgments }

We would like to thank S.~Martin, A.~Pierce, K.~Tobe and J.~Wacker for
helpful conversations. We also wish to acknowledge the collaboration with the authors of
\cite{Gambino:2005eh} in comparing our results with theirs, which resulted in this revised version.
We also thank $DOE$ and $MCTP$ for support.


\end{document}